\documentclass[acmsmall]{acmart}

\acmJournal{PACMPL}
\acmVolume{1}
\acmNumber{HATRA} 
\acmArticle{1}
\acmYear{2022}
\acmMonth{12}
\acmDOI{} 
\startPage{1}

\setcopyright{none}

\usepackage{booktabs}   
\usepackage{subcaption} 

\usepackage{comment}
\usepackage{enumerate}
\usepackage{amsmath,graphicx,algorithmic,xspace,url}
\usepackage[titlenotnumbered,noend,noline]{algorithm2e}

\usepackage{array}
\usepackage{listings}
\usepackage{xpatch,letltxmacro}
\LetLtxMacro{\cminted}{\minted}

\xpretocmd{\cminted}{\RecustomVerbatimEnvironment{Verbatim}{BVerbatim}{}}{}{}

\usepackage{xcolor}
\usepackage{listings}

\begin{document}

\title{Exploring the Verifiability of Code Generated by GitHub Copilot}


\author{Dakota Wong}
\affiliation{
  \institution{University of Waterloo}            
  \country{Canada}                    
}
\email{d82wong@uwaterloo.ca}          

\author{Austin Kothig}
\affiliation{
  \institution{University of Waterloo}            
  \country{Canada}                    
}
\email{akothig@uwaterloo.ca}          

\author{Patrick Lam}
\orcid{0000-0001-8278-5400}             
\affiliation{
  \institution{University of Waterloo}            
  \country{Canada}                    
}
\email{patrick.lam@uwaterloo.ca}          


\begin{abstract}
GitHub's Copilot generates code quickly. We investigate whether it generates good code. Our approach is to identify a set of problems, ask Copilot to generate solutions, and attempt to formally verify these solutions with Dafny. Our formal verification is with respect to hand-crafted specifications. We have carried out this process on 6 problems and succeeded in formally verifying 4 of the created solutions. We found evidence which corroborates the current consensus in the literature: Copilot is a powerful tool; however, it should not be ``flying the plane" by itself.

\end{abstract}



\maketitle

\newcommand{\ie}{\textit{i.e.,} }
\newcommand{\cf}{\textit{cf.} }
\newcommand{\eg}{\textit{e.g.,} }
\newcommand{\etal}{\textit{et al.}}

\newcommand{\gil}[1]{{\textcolor{blue}{\textbf{Gil: #1}}}}
\newcommand{\jeehoon}[1]{{\textcolor{green!60!black}{\textbf{Jeehoon: #1}}}}
\newcommand{\yoonseung}[1]{{\textcolor{pink!40!black}{\textbf{Yoonseung: #1}}}}
\newcommand{\youngju}[1]{{\textcolor{pink!60!black}{\textbf{YoungJu: #1}}}}
\newcommand{\juneyoung}[1]{{\textcolor{brown}{\textbf{Juneyoung: #1}}}}
\newcommand{\sanghoon}[1]{{\textcolor{brown!60!black}{\textbf{Sanghoon: #1}}}}
\newcommand{\dongyeon}[1]{{\textcolor{brown!40!black}{\textbf{Dongyeon: #1}}}}


\definecolor{keywords}{RGB}{199, 21, 133}
\definecolor{comments}{RGB}{34,139,34}
\definecolor{variables}{RGB}{0, 0, 255}
\definecolor{black-background}{RGB}{0, 0, 0}
\definecolor{main-color}{RGB}{255, 255, 255}
\definecolor{cream}{RGB}{255,253,208}

\lstdefinelanguage{Dafny}
    {sensitive = false,
    keywords = {},
    otherkeywords = {<,>,<=,>=,:=,==,!=,<==,==>,<==>,+,-,*},
    keywords = [2]{while, invariant, forall, exists, predicate, method, lemma, returns, return, ensures, requires, if, else, reads, modifies, calc, array, assert, assume, bool, int, decreases, function, ghost, import, in, include, new, null, object, reads, seq, set, string, then, this, true, false, var},
    morekeywords = [3]{binarysearch, largest_sum, sortArray, primeFactors, maxHeapify, maxHeaped, isr, append_seq_mul, multSeq, pivot, multiset, sorted, Sum_Array, max_sum_subarray, found, not_found, distinct, old},
    keywordstyle = {\color{teal}},
    keywordstyle = [2]{\bfseries\color{keywords}},
    keywordstyle = [3]{\bfseries\color{blue}},
    numberstyle = {\color{black}},
    morecomment = [l][\color{comments}]{//},
    basicstyle = {\footnotesize\ttfamily \color{black}},
    breaklines = true,
    keepspaces=true,                 
    numbers=left,                    
    numbersep=10pt,
    }

\lstnewenvironment{Dafny_env}
    {\lstset
        {language=Dafny,
        frame=shadowbox,
        framexleftmargin = 10mm,
        framexrightmargin = 10mm,
        framextopmargin = 3mm,
        framexbottommargin = 3mm,
        rulesepcolor = \color{gray},
        backgroundcolor = \color{cream}
        }
    }
    {}


\lstdefinelanguage{py_custom}
    {sensitive = false,
    keywords = {},
    otherkeywords = {<, >, =, ==, !=, +, -, *, /, \%},
    keywords = [2]{def, list, int, while, var, if, elif, else, for, return, in},
    morekeywords = [3]{binary_search, largest_sum, sortArray, primeFactors, maxHeapify, len},
    keywordstyle = {\color{teal}},
    keywordstyle = [2]{\bfseries\color{keywords}},
    keywordstyle = [3]{\bfseries\color{blue}},
    numberstyle = {\color{black}},
    morecomment = [l][\color{comments}]{\#},
    morecomment = [s][\color{comments}]{"""}{"""},
    basicstyle = {\footnotesize\ttfamily \color{black}},
    breaklines = true,
    keepspaces=true,                 
    numbers=left,                    
    numbersep=10pt,
    }

\lstnewenvironment{python_env}
    {\lstset
        {language=py_custom,
        frame=shadowbox,
        framexleftmargin = 10mm,
        framexrightmargin = 10mm,
        framextopmargin = 3mm,
        framexbottommargin = 3mm,
        rulesepcolor = \color{gray},
        backgroundcolor = \color{cream}
        }
    }
    {}

\newcommand{\etc}{\textit{etc.}}
\newcommand{\code}{\texttt}
\def\changemargin#1#2{\list{}{\rightmargin#2\leftmargin#1}\item[]}
\let\endchangemargin=\endlist 



\section{Introduction} \label{sec:Introduction}

In aviation, commercial airplanes have two yokes (controls) which allows for sharing of responsibilities and workload between a primary pilot and a copilot. Both operators are fully trained, informed, and capable of flying the aircraft; however, only one of them needs to be controlling the aircraft at any one moment. In software development, traditionally, code is written by an individual. In team environments, this code is typically scrutinized and assessed by a number of people, but originally it had to be developed by an individual. Similar to a pilot, software developers are ordinarily well trained, informed, and capable of programming; however, inefficient or incorrect code segments can potentially slip between the cracks and into code-bases when a developer does not fully understand their tools, the domain, or the required program specifications.

GitHub, with the use of OpenAI's Codex~\cite{chen2021evaluating}, has released an ``AI pair programmer'' to assist in software development~\cite{github2022copilot}. This programming partner is aptly named Copilot. The aim of this tool is to shift the programmer's time and attention away from boilerplate and repetitive code patterns, and toward the critical design of the system. Copilot works by providing \textit{implementation suggestions} based on natural language descriptions of the required code. Copilot is currently capable of generating code in eleven programming languages, including C, C++, C\#, Go, Java, JavaScript, PHP, Python, Ruby, Scala, and TypeScript\footnote{\url{https://docs.github.com/en/copilot/overview-of-github-copilot/about-github-copilot}}.

The community is struggling to understand what Copilot can and cannot do, and proofs are one indicator of where Copilot might stand in terms of generating correct code. We know that Copilot definitely generates some code, but what can we conclude about the code that it generates? Thus, our goal is to gain some understanding of the quality of Copilot's generated code; our approach is to attempt to verify selected examples of Copilot-generated code using the software verification language Dafny~\cite{leino2010dafny}.


\section{Related Work} \label{sec:RelatedWork}

There has been a great deal of scrutiny of Copilot and of the code it produces. Early studies which aim to empirically evaluate Copilot~\cite{nguyen2022empirical} have found that while this early version of the model is not perfect, its current results are impressive. In their work, \citeANP{nguyen2022empirical} tested Copilot against $33$ LeetCode\footnote{\url{https://leetcode.com/}} questions ($4$ from the easy category, $17$ from the medium category, and $12$ from the hard category) using four programming languages (Python, JavaScript, Java, C), ran the generated code against the LeetCode test cases, and examined the percentage of passing cases. Java performed best, being able to solve all test cases for $19$ of the $33$ problems. The authors found that, often, the code that Copilot generated could be simplified, and that sometimes the generated code would require methods or packages that have not been defined or declared. 

Copilot generates code differently from traditional genetic programming paradigms. \citeN{sobania2022choose} found that Copilot could generate correct solutions to a variety of problems, with comparable accuracy to genetic programming. The improvement that Copilot makes over traditional approaches is that it was trained on a $159$ GB dataset of unlabelled data, whereas genetic programming requires the dataset to be rigorously labeled. Thus, \citeANP{sobania2022choose} claimed that the code Copilot generates is often easier to read and understand compared to that generated by genetic programming.

It is clear from the current research that Copilot is not a magic wand solution to problems in software engineering. With its flaws, how are developers able to benefit from this tool effectively? \citeN{dakhel2022github} compared Copilot's proposed solutions against human programmers on a set of fundamental problems. 
The authors found that Copilot was unable to understand certain limitations embedded in the problem description, and while understanding the fundamentals of what is being requested, it can sometimes create errors when it does not understand. They conclude that while the generated code from Copilot is not a perfect solution, it can still be incorporated and modified into a project for faster early-stage prototyping.

Continuing on the question of usability of Copilot-generated code by developers, 
\citeN{vaithilingam2022expectation} performed a user study. They found that
using Copilot did not improve the task completion time or success rate, but that
the developers preferred to start from the Copilot-generated code rather than a 
blank slate. The researchers also pointed out that it was challenging for 
developers to work with long blocks of generated code (and indeed, some developers
failed to complete on time due to difficulties debugging generated code)---attempting 
to verify the code would also be a challenge, though one that might provide more 
understanding. One difference between our study and theirs, though, is that
we did not encounter generated solutions that were known incorrect, and their
users did.

Any introduction of small errors can potentially compound over time, creating large vulnerabilities in a software stack.
\citeN{pearce2022asleep} investigated code generated by Copilot for known security vulnerabilities. The authors prompted Copilot to generate code for $89$ cybersecurity scenarios. They found that approximately $40\%$ of generated code snippets were vulnerable to some form of known attack. Also, \citeN{siddiq22:_empir_study_code_smell_trans} found code smells (including security smells) in training sets used for transformer-based code generation models for Python as well as in suggestions produced by Copilot. If code generated by Copilot is to be used in production, it needs to be rigorously vetted for exploits (and keeping in mind the caveat from \citeANP{vaithilingam2022expectation} about understanding Copilot's generated code).


\section{Methodology} \label{sec:Methodology}


Our goal is to apply verification techniques to code generated by GitHub Copilot. We came up with sample problems and tasked Copilot with generating solutions in Python. We then used Dafny to verify that Copilot's generated code satisfies the requirements.

More specifically, our methodology is as follows. First, we chose 6 problems to give to Copilot (five that we chose ourselves and one from \citeN{nguyen2022empirical}). We wanted to test whether Copilot could generate a function which would solve the problems; by ``solve'', we mean that the generated code satisfies the problem's formal specifications. Thus, we wrote formal specifications in Dafny for each problem, consisting of \textit{requires} and \textit{ensures} statements.

Having defined the problems, we then asked Copilot to generate solutions in Python. To give Copilot the best chance of generating a good solution, we gave it a full function signature, including a descriptive function name, parameter name(s), parameter type(s), and an output type. With this input, Copilot then generated the function implementation. Since our goal was to test the capabilities of Copilot, we always chose the first suggested function.

Next, we carefully translated each Copilot implementation from Python to Dafny and added the specifications that we had created. We took care to preserve the behaviour of the implementation. In all cases, the implementations contained a loop of some type. Hence, the provided \textit{requires} and \textit{ensures} statements were inadequate to generate a proof that verified the implementation. We thus attempted to manually derive loop invariants so that Dafny could verify the implementation. During the verification process, we added extra \textit{requires} and \textit{ensures} clauses, if they were needed to verify the program, along with additional helper functions and methods when necessary.



\section{Results} \label{sec:Results}

We have attempted to verify 6 Copilot implementations, one of which intersects the LeetCode problems in \cite{nguyen2022empirical}. Table~\ref{tab:stats} summarizes the algorithms, implementations, and verification attempts. Incidentally, during testing, consistent with \citeN{nguyen2022empirical}, we found that Copilot would periodically include packages or methods that had not been defined; it happened to not do so for any of the implementations we describe here.

We have archived the current version of the Copilot-generated implementations and our Dafny code at \url{https://zenodo.org/record/7040924}.

\begin{table}[h]
    \centering
    \caption{Algorithm implementations generated by Copilot. Dafny verification attempt statistics.}
    \begin{tabular}{  l r l c   rr  } 
   \textbf{Algorithm} & \textbf{Python}  & \textbf{Verification} &  \textbf{Verified} & \textbf{Specification} & \textbf{\# Invariants} \\
  \textbf{Name} & \textbf{LOC} & \textbf{ Difficulty} & & \textbf{LOC} &  \\
  \hline
  Binary Search (\ref{sub:binary_search}) & 12 & Easy   & $\checkmark$ & 5 & 3 \\
  Two Sums      (\ref{sub:two_sums})      &  5 & Medium & $\checkmark$ & 6 & 7 \\
  Largest Sum   & 10 & Medium & $\checkmark$ & 2 & 3 \\
  Sort Array    &  6 & Medium & $\checkmark$ & 4 & 11 \\
  Prime Factors (\ref{sub:prime_factors}) & 12 & Hard   &              & 3 & $>$ 23 \\
  Max Heapify   (\ref{sub:max_heapify})   & 14 & Hard   &              & 4 & $>$ 4 \\
\end{tabular}
    \label{tab:stats}
\end{table}

Figure~\ref{fig:prompts} presents the Copilot prompts that we used to generate implementations.

\begin{figure}[h]
\begin{center}
\begin{verbatim}
  def binary_search(arr: list, target: int) -> int:
  def twoSum(self, nums, target): // LeetCode; types on omitted subsequent lines
  def largest_sum(nums: list) -> int:
  def sortArray(arr: list) -> list:
  def primeFactors(val: int) -> list:
  def maxHeapify(arr: list) -> list:
\end{verbatim}
\end{center}
\caption{Copilot prompts for our algorithms.\label{fig:prompts}}

\end{figure}

We were able to create a valid proof in Dafny for 4 of the 6 solutions generated by Copilot; we next discuss 2 of the valid proofs and our 2 failed attempts.

\subsection{Binary Search} \label{sub:binary_search}

``Binary Search'' is an algorithm which searches a sorted array for the index of a target element. The algorithm is appealing as it has a worst-case linear time complexity $O(n)$. Implementations are recognizable and simple to understand, making this algorithm a good candidate for our first verification example.

This example was inspired by a Dafny guide written by Arie Gurfinkel\footnote{\url{https://ece.uwaterloo.ca/~agurfink/stqam/rise4fun-Dafny}}. Our first observation was that Copilot's implementation of Binary Search is different from that in the Dafny guide, leading to a different verification process. 

We constructed formal specifications for binary search. As input, the implementation will take an array of distinct sorted integers and an integer target. It should then search the array for the target, returning the index of the target if found. If the target is not found, the program should have a return value (\textit{r.v}) of $-1$. These specifications can be expressed as:

\vspace{-12px}
\begin{align*}
    &\textbf{requires } \; \mathrm{Sorted}(arr)\\
    &\textbf{requires } \; \mathrm{Distinct}(arr)\\
    &\textbf{ensures } \;\;\; r.v \neq -1 \implies \mathrm{found}(arr, \; target, \; r.v)\\
    &\textbf{ensures } \;\;\; r.v = -1 \implies \mathrm{not\_found}(arr, \; target)
\end{align*}
\vspace{5px}

In the specifications above, \emph{Sorted}, \emph{Distinct}, \emph{found}, and \emph{not\_found} are all predicates. \emph{Sorted} checks whether an array is sorted in ascending order and \emph{Distinct} checks whether all elements of an array are distinct. We require that valid inputs satisfy these predicates. Next, we have the predicate \emph{found}, which checks to see if the target was found at the index returned by the algorithm. This predicate should be true when the returned index is not $-1$. To handle the case where the target is not found, we have predicate \emph{not\_found} that checks that the target does not exist in the array. This predicate is implied to be true when the return value is $-1$.

In parallel with manually producing specifications, we used Copilot to generate an implementation, given the following signature:

\vspace{10px}
\begin{python_env}
def binary_search(arr: list, target: int) -> int:
\end{python_env}
\vspace{10px}

\newpage 
We accepted the first suggested implementation by Copilot based on the provided input:

\begin{python_env}
def binary_search(arr: list, target: int) -> int:
    low = 0
    high = len(arr) - 1
    while low <= high:
        mid = (low + high) // 2
        if arr[mid] == target:
            return mid
        elif arr[mid] < target:
            low = mid + 1
        else:
            high = mid - 1
    return -1
\end{python_env}

As stated above, we manually translated the implementation to Dafny, and added the formal specifications that we came up with. This also involved defining the predicates \emph{Sorted} and \emph{Distinct}. Given a specification and implementation, we could start verification.

However, Dafny also requires loop invariants. To provide them, we needed to (manually) reason about what happens in the implementation's while loop. From our familiarity with binary search, we know that it uses two indices, high and low. High starts at the last element of the array; low starts at the first. In each iteration, the indices move closer together, closing the window of elements where the target can exist. In particular, we know that the target cannot exist at indices less than or equal to that of the low pointer or at indices greater than equal to that of the high pointer. This logic was to implement the primarily loop invariant of the algorithm. In first-order logic,

\vspace{-10px}
\begin{align*}
    &\forall i \cdot 0 \leq i \leq \textit{low} \land \textit{high} \leq i < \textit{arr.Length} \implies \textit{arr}[i] \neq \textit{target}.
\end{align*}

With the loop invariants, Dafny was able to verify the implementation. So, in this case, Copilot was able to successfully generate an implementation which satisfied our formal specifications for binary search. This implementation could be verified without adding any extra \textit{requires} or \textit{ensures} statements beyond what we had initially identified as the specification. 

\subsection{Two Sums} \label{sub:two_sums} 

We next picked a problem from LeetCode that was used in the study by \citeN{nguyen2022empirical}. ``Two sums'' takes an array of integers and an integer target, and returns two distinct indices in the array where the elements at those indices sum to the target. The Copilot input was:

\vspace{10px}
\begin{python_env}
    def twoSum(self, nums, target):
        """
        :type nums: List[int]
        :type target: int
        :rtype: List[int]
        """
\end{python_env}
\vspace{10px}
The solution that Copilot generated passed all of the LeetCode test cases. LeetCode provides a number of constraints on the inputs---notably that a valid answer always exists. We used the constraints to formulate the following requires and ensures clauses:

\vspace{-12px}
\begin{align*}
    &\textbf{requires } \; 2 \leq \mathrm{nums.Length}\\
    &\textbf{requires } \; \exists i, j \cdot (0 \leq i < j < \code{nums.Length} \wedge \code{nums}[i] + \code{nums}[j] == \code{target})\\
    &\textbf{ensures } \;\;\; \code{index1} \neq \code{index2}\\
    &\textbf{ensures } \;\;\; 0 \leq \code{index1} < \code{nums.Length}\\
    &\textbf{ensures } \;\;\; 0 \leq \code{index2} < \code{nums.Length}\\
    &\textbf{ensures } \;\;\; \code{nums}[\code{index1}] + \code{nums}[\code{index2}] == \code{target}
\end{align*}

We converted the Python to Dafny and, with moderate effort, we then
managed to write invariants that enabled Dafny to verify the generated
solution. The 5 lines of Python cost resulted in 8 lines of Dafny.
We added 7 invariants (3 in the outer loop and 4 in the inner loop) to enable
the verification.

Hence, in this case, Copilot generated a solution that passes the LeetCode test
cases and that we managed to verify.

\subsection{Prime Factors} \label{sub:prime_factors}

``Prime Factors'' is an algorithm that takes an integer and returns a list of prime numbers, which, when multiplied together, reproduce the original number. For example, the number $18$ decomposes into the list $[2, 3, 3]$. The algorithm is $O(n)$, but does contain a nested loop.


We formally specified this algorithm; it accepts an integer greater than or equal to $2$, and returns a list of prime factors of the integer. Our (arithmetic-heavy) specifications were:

\vspace{-10px}
\begin{align*}
    &\textbf{requires} \;\; \code{val} \geq 2\\
    &\textbf{ensures}  \;\; \prod_{\code{i}=0}^{\mathrm{len}(\code{arr})} \code{arr[i]} == \code{val}
\end{align*}


With these simple formal specifications defined, we were ready to prompt Copilot to generate our algorithm. We provided Copilot with the following function signature:

\vspace{10px}
\begin{python_env}
def primeFactors(val: int) -> list:
\end{python_env}
\vspace{8px}

\noindent
which prompted Copilot to suggest the following Python implementation:

\vspace{10px}
\begin{python_env}
def primeFactors(val: int) -> list:
    factors = []
    while val 
        factors.append(2)
        val = val // 2
    for i in range(3, int(val ** 0.5) + 1, 2):
        while val 
            factors.append(i)
            val = val // i
    if val > 2:
        factors.append(val)
    return factors
\end{python_env}
\vspace{10px}

To convert to Dafny: 1) As with previous implementations, we adapted Python \code{for} loops into \code{while} loops. 2) We had to implement integer square root in Dafny. Verifying integer square root is nontrivial; however, we found a verified implementation\footnote{\url{https://homepage.cs.uiowa.edu/~tinelli/classes/181/Spring11/Tools/Dafny/Examples/square-root.dfy}} which we used as a library. 3) We had to choose a Dafny data structure equivalent to a dynamic array.

For dynamic arrays (the Python equivalent of \code{list}),
we used the Dafny \code{seq<int>} (sequence of integers), as it had a built in concatenate method which acts by \textit{adding} to sequences to produce a new sequence which can be assigned. However, initialization was a challenge: Dafny does not permit empty sequences. We thus initialized our sequence with multiplicative unit 1; though this is not identical to the Python solution, it is reasonably equivalent.

We also created a function for multiplying the elements of the sequence, adding invariants to check that the current product times the remaining value produces the original value.

Using our chosen invariants, we encountered Dafny timeout issues and failed to verify the implementation. It may be the case that our invariants are correct and Dafny is unable to verify them in a reasonable timeframe, but we don't know that. As a last resort, to try to help the proof resolve with ``what we know to be true", we added \textit{lemma}s which assumed false, allowing us to say two specific \textit{ensures} clauses were true:

\vspace{-10px}
\begin{align*}
    & \textbf{ensures} \;
        \left(\prod_{\code{i}=0}^{\mathrm{len}(\code{arr})} \code{arr[i]}\right) \cdot v == \code{val}\\
    & \textbf{ensures} \;
        (v == 1) \implies \left(\prod_{\code{i}=0}^{\mathrm{len}(\code{arr})} \code{arr[i]}\right) == \code{val}
\end{align*}

\noindent
before and after each \textit{append} of a new value to the list of factors. Even assuming these two clauses to be true, the Dafny solver continued to hit the time limit for finding a proof. Thus, although we believe that Copilot generated a correct implementation, we have not yet been able to verify it.

\subsection{Max Heapify} \label{sub:max_heapify}

``Max Heapify'' is an algorithm for turning an array into a max-heap data structure. The array is conceptually a complete binary tree, where each index holds the value of a particular node, and that node's left and right children's indices are found using the following mapping:

\vspace{-10px}
\begin{align*}
    \code{left\_child}  &= (\code{index} * 2) + 1\\
    \code{right\_child} &= (\code{index} * 2) + 2.
\end{align*}
\noindent
A max-heap satisfies the property that each node is greater than or equal to its left and right child. Since this property is recursive, it implies that any particular node is (transitively) larger than every node in both its right and left children, so the largest element in the tree should be at the root.

With this in mind, we wrote formal specifications:

\vspace{-10px}
\begin{align*}
    &\textbf{requires} \;\; \code{arr} \; \text{is a} \; \code{list} \; \text{of distinct integers}\\
    &\textbf{ensures} \;\;\; \forall x \cdot (0 \leq x < \mathrm{len}(\code{arr})) \implies \\
    &\quad\quad (((2x + 1 < \mathrm{len}(\code{arr})) \implies (\code{arr[} x \code{]} \geq \code{a[} 2x + 1 \code{]})) \; \land \\
    &\quad\quad \;\; ((2x + 2 < \mathrm{len}(\code{arr})) \implies (\code{arr[} x \code{]} \geq \code{a[} 2x + 2 \code{]}))).
\end{align*}

\noindent
That is, the input array's values are unique; and the resulting list has the properties of a max-heap.
Simultaneously, we provided the following function signature to Copilot:

\vspace{10px}
\begin{python_env}
def maxHeapify(arr: list) -> list:
\end{python_env}
\vspace{8px}

Copilot produced a 14-line Python implementation, which we translated into Dafny and attempted to verify. We designed a predicate based on the \textit{ensures} clause, which checks that each node is greater than or equal to the node at its childrens' indices, if those indices are not out of bounds for the array.

We noticed that the Copilot implementation differs from more traditional implementations by sifting down rather than sifting up: the recursive call to \code{maxHeapify} is given the entire (mutated) array. Often, implementations of max-heap traverse all non-leaf nodes backwards to the root and sift-up (with a recursive call) starting at any nodes that are out of place.


Prior to formal Dafny verification, we tested the the generated algorithm on a series of random inputs to ensure it produced a max-heap each time. Inspecting the algorithm with Python debugging tools showed that the trace-tree for this implementation goes quite deep. The implementation appears to make recursive calls, starting from the root, enough times to ensure that eventually the max-heap property holds. This would make finding a verifiable proof difficult. Instead of the standard \textit{heapify} algorithm running in $O(n \; log \; n)$, the version Copilot generated for us appears to run in time $O(n^2)$. 

We tried a number of different invariants to show that the array was \textit{partially} heapified for each step in the loop. However, we were unable to prove termination of the loop to begin with. For this reason, we were unable to formally verify the generated algorithm. 

\section{Discussion} \label{sec:Discussion}

Copilot is undoubtedly good at generating code. In this work, we took a formal verification approach to investigating the code that Copilot generates, using Dafny. Of the 6 examples we tried, we managed to verify the 4 simpler problems. One of the problems appeared in~\cite{nguyen2022empirical}, which used LeetCode test suites to establish correctness of Copilot-generated code.

Our work produced corroborating evidence that Copilot's generated code might be dubious. It also explored the difficulty of using Dafny to verify implementations; although Dafny is powerful, it can be nontrivial to verify implementations using it.

We believe that our work reinforces the fact that to effectively deploy Copilot, one should know what is needed. Similar to the work by \citeN{dakhel2022github}, we found that Copilot appears to skip over certain constraints imposed on the input space. As a specific example, the ``two sums'' example assumes that a solution exists; we had to manually record that assumption in our specification. Humans are not good at reasoning about edge cases~\cite{wheeler21:_apple}, and Copilot reproduces that bias. Thus, to productively use Copilot, it remains incredibly important to reason carefully about what the problem specifications should be.


We also found that verifying the generated software with Dafny was somewhat difficult, both due to Dafny implementation language limitations, and---more importantly---verifiability reasons. Some Python idioms do not translate cleanly to Dafny; we had to reformulate Python generator objects (\eg \code{range} in a \code{for} loop) as \code{while} loops. Other Python data types such as \code{list} don't have an obvious substitution in Dafny. Dafny also omits math operations such as floor and square root, which Copilot generated in its Python solutions.

In the LeetCode study~\cite{nguyen2022empirical}, not all of the Copilot-generated solutions passed the LeetCode test cases. In this work, we did not attempt to verify any solutions that failed their test cases. Of course, we expect such solutions to be unverifiable. However, attempting verification could provide another way to understand how the solutions are incorrect, beyond simply inspecting the generated code. As is true of verification in general, an attempted verification can provide insight into the code, complementing a test-based lens.

We believe that algorithms designed with verifiability in mind are more likely to be easier to prove. A proof of correctness of binary search halves the search space at each step. If an implementation follows suit, we can focus on solving a subset of the problem in its proof. In reasoning about a sorting algorithm, we would like to be able to show that each step of a loop results in the array being partially sorted. 

When we were choosing algorithms to verify, we thought that max-heap would be simple, as it has very clear sub-problems which, when solved, result in a well-defined data structure. However, the Copilot-generated implementation does not obviously reduce to sub-problems. We believe that an iterative version of the max-heap algorithm which walks backwards through the tree would have been easier to verify than the recursive version which Copilot generated for us. 

Of course, we have not established that proving the generated max-heap implementation is impossible. Per \citeN{leino2010dafny}, Dafny verification often requires an expert, and it is not an easy task for the average programmer. \citeN{leino2014dafny} have also identified that, for some problems, the solver timing out can be a barrier to verifiability. There are some shortcuts which can be taken in the proof to avoid timing out; however, these shortcuts require a great amount of practice to identify. Max-heapify has been formally verified~\cite{tafat11:_binar_why3} using other software verification techniques, but the proof and its description are both reasonably lengthy.







\paragraph{Generalizability}
In this work, we attempted to verify 6 specific implementations generated by Copilot against specifications that we manually constructed. We found that we were able to verify 4 of the 6 implementations. The remaining 2 generated implementations may be correct, and others may be able to verify them, but we were not. We've thus shown that it is possible to verify some Copilot-generated code, and that other code is at least nontrivial to verify. We also know from~\cite{nguyen2022empirical} that, provided with a textual prompt and test cases, Copilot can generate code that does not output the expected results.

Our Python-mediated translation process meant that we avoided Dafny's treatment of programs with pointers and pointer arithmetic. Because we asked Copilot for Python implementations, and these implementations do not include pointer and memory management, we could not explore how Dafny can be used to verify functional correctness for pointer-based programs~\cite{leino2010dafny}.

We speculate that, even if we had chosen different target languages and verification tools, the fundamental difficulty of finding invariants would still be an issue with verifying code from Copilot. The fact that Copilot generated a recursive implementation that we could not verify means that we cannot blame Copilot for generating difficult-to-verify loops. Dafny's lack of square root implementation is, of course, a Dafny-specific problem.

Of course, our results would be stronger with more verification attempts, and across different domains as well. LeetCode examples, in particular, are more likely than average to be represented in the Copilot training corpus. Going beyond algorithmic examples (e.g. reading a CSV file, connecting to a database) would also be interesting, though harder to formally specify.

The other source of variability is expertise in using verification tools. We are not experts in using these tools, and hence our experience may be representative of average users, but the sample size of 2 verification users is small. A study involving more developers attempting to verify code would provide stronger results.

\paragraph{Concluding thoughts}
In our experience, Copilot is a powerful tool for quickly producing (prototype) implementations: it can generate code faster than one could possibly write it. However, Copilot is certainly not a magic wand. A programmer using Copilot would be well advised to not only look at what Copilot has produced, but also ensure it is correct and appropriate for the intended use case. If not with formal verification, one ought to at least rigorously vet the generated code with a test suite, prior to integrating it into any system. One should also consider copyright issues~\cite{howard2021github}. Of course, in principle, one should vet any code before integrating into a system, whether human produced or machine produced.

We observed that Copilot was quite capable when generating solutions to well known or established problems, likely related to the fact that Copilot's training data came from user data on GitHub. Like~\citeN{dakhel2022github}, we found that Copilot generated sub-optimal and possibly incorrect solutions to some of our problems, \eg Max Heapify. Due to how Copilot is trained, we do not believe that it would work as well on esoteric problems; it would be interesting to explore that point.


\bibliography{references}


\end{document}